# Genetic optimization "Invents" non-Hermitian potentials for Asymmetric Reflectivity

W. W. Ahmed[1], R. Herrero[2], M. Botey[2], Y. Wu[1*], and K. Staliunas[2,3,4†]

[1]Division of Computer, Electrical and Mathematical Sciences and Engineering, King Abdullah University of Science and Technology (KAUST), Thuwal, 23955-6900, Saudi Arabia

[2]Departament de Física, Universitat Politècnica de Catalunya (UPC), Colom 11, E-08222 Terrassa, Barcelona, Spain

[3]Institució Catalana de Recercai Estudis Avançats (ICREA), Passeig Lluís Companys 23, E-08010, Barcelona, Spain

[4]Vilnius University, Laser Research Center, Saulėtekio al. 10, Vilnius, Lithuania

Emails: [*]ying.wu@kaust.edu.sa, [†]kestutis.staliunas@icrea.cat

## Abstract

We propose a general design strategy based on genetic optimization to realize asymmetric reflectivity in periodic and non-periodic planar structures containing dielectric and gain-loss layers. By means of an optimization algorithm it is possible to design the imaginary (or real) part of the complex permittivity distribution from any given and arbitrary real (or imaginary) permittivity distribution, i.e to create non-Hermitian potentials intended to achieve "*on demand*" light transport for a selected spectral range. Indeed, the asymmetric response of the obtained complex permittivity distribution is directly related to its area in the complex permittivity plane. In particular, unidirectional light reflection can be designed in such a way that it switches from left to right (or vice versa) depending on the operating frequency. Moreover, such controllable unidirectional reflectivity can be realized using a stack of dielectric layers while keeping the refractive index and gain-loss within realistic values.

## Introduction

The transport of light is symmetric with respect to the propagation direction in conservative, Hermitian systems, as guaranteed by reciprocity and energy conservation principles. However, unidirectional light transport devices, preventing back reflections, would be highly desirable in integrated photonics to design a new generation of chip-scale optical devices for different technological applications[1-3]. As demonstrated in recent years, such breaking of wave propagation symmetry is possible in materials with complex permittivity profiles, i.e,. with non-Hermitian (in particular PT-symmetric) potentials. Non-Hermitian optics are attracting increasing attention particularly since reported in photonics due to several spectacular features, including unidirectional invisibility[4,5], optical switching[6], coherent perfect absorption[7,8], beam refraction[9], nonreciprocity of light propagation[10-12], nonlinear effects[13-15], among others[16]. In optics, the general requirement for PT-symmetry is that the complex permittivity, obeys the condition: $\varepsilon(r) = \varepsilon^*(-r)$ i.e., the real part of the permittivity distribution is a symmetric function while the imaginary part is an antisymmetric function in space, while this condition may be generalized to include anti-PT-symmetry or a wider class of non-Hermitian potentials[16-18], including stacked layers of purely dielectric and purely magnetic slabs[19]. The physical mechanism behind totally asymmetric light propagation is a sharp symmetry breaking-transition when a non-Hermiticity parameter exceeds a certain critical value, which is referred as exceptional point[9,10]. While non-Hermitian potentials may still hold real eigenvalues, in the broken phase regime the eigenvalues of the systems become complex.

To date, several realizations of non-Hermitian media have been explored to exploit an asymmetric light propagating for extraordinary functionalities[20-22]. However, simple planar non-Hermitian structures comprising of a stack of dielectric layers with gain and loss are of particular interest for their compatibility with the existing fabrication technologies and availability of their theoretical models for deep physical insights in scattering mechanisms. The specific asymmetric behavior of non-Hermitian structures can be achieved by carefully tuning the material constituents and geometry parameters. The choice of the structural and optical parameters is usually based on intuition through trial-and-error methods. However, with the continuous increase in performance and integration level demands, the tedious trial and error methods become inefficient and severely time consuming. In this scenario, optimization approaches have stirred the development of on-demand functionalities and integration level of optical devices. The heuristic optimization techniques, relying on evolutionary algorithms, such as the particle swarm optimization (PSO)[23,24] and genetic algorithm (GA)[25-26] are widely used to solve single and multi-objective problems. However, GA is proved to be very efficient in searching the global maxima for complex optimization problems and high-performance designs[27-30]. In this article, we propose a feasible approach based on genetic optimization to design multifrequency reflectionless periodic and non-periodic planar structures. We combine GA with the general Transfer Matrix Method (TMM) to optimize the distribution of complex permittivity in multilayered structures for the desired manipulation of the propagating light, either in broadband or in frequency-selective operation.

To design a simple unidirectional reflectionless system, we set, as a target function, different (say 0% and 100%) reflectivities illuminating either from the right or from the left, yet the same 100% symmetric transmission. In addition, we allow the index and the gain/losses to vary only in particular limits. For simplicity, we take a 1D system divided into several spatial domains, with different values of the complex permittivity, expecting that the optimization algorithm brings the system to some PT-like structure [see Fig.1]. The idea is to seed the real part (or alternatively the imaginary part) of the complex permittivity in each section of the 1D stack and let the algorithm to choose the corresponding counterpart of permittivity for unidirectional behavior. This article aims at designing periodic and non-periodic non-Hermitian structures, from a given arbitrary permittivity profile, that hold asymmetric light propagation at different spatial frequencies, as illustrated in Fig. 1a. Figures 1b and 1c, depict the asymmetric spectra corresponding to the periodic structures with discrete unidirectional reflection at two given frequencies, either in the same or opposite directions. In Fig. 1d, we intend to design a non-periodic planar structure that exhibits broadband unidirectional propagation as it follows from the generalized Hilbert transform[21].

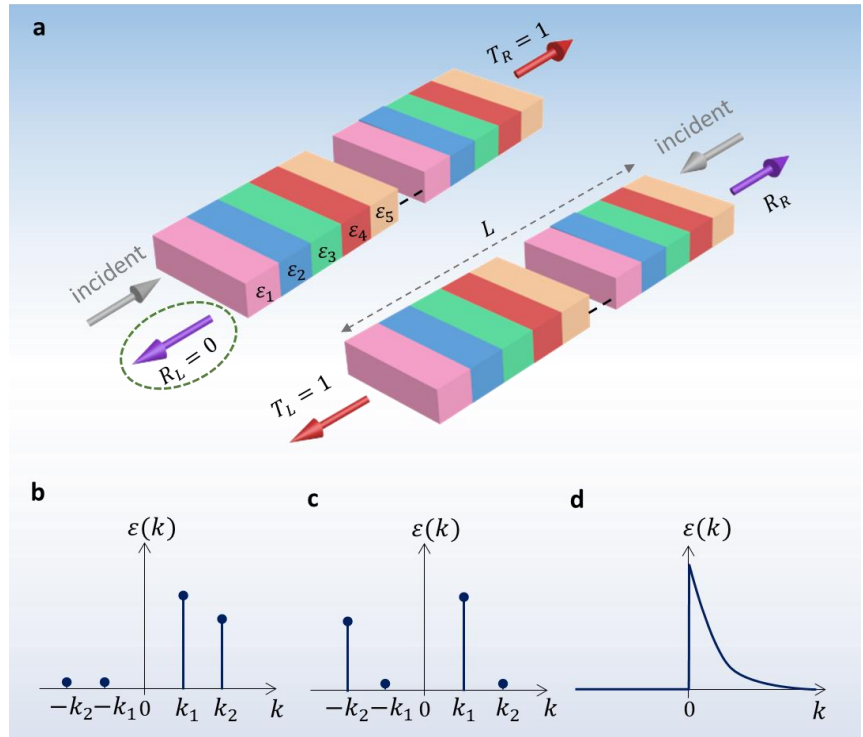

**Fig.1** Design principle for frequency selective and broadband asymmetric propagation. **a** Schematic design of a planar PT-symmetric like periodic structure for asymmetric left/right reflection. **b-c** Asymmetric spectra holding unidirectional reflection at two discrete frequencies (due to periodicity), either in the same or opposite directions, in periodic non-Hermitian structure. **d** Non-periodic structure showing a broad-spectra unidirectional reflection.

# Results

To realize '*on demand*' asymmetric reflectivity, we consider an optical structure consisting on a stack of 1D layers with complex dielectric permittivity indexed as $\varepsilon^j = \varepsilon_r^j + i\varepsilon_i^j$, corresponding to the *j*th layer [see Fig. 1]. The structure is embedded in a homogeneous medium with uniform permittivity $\varepsilon_0$. We use the TMM to obtain the spectral transmission $(T_R, T_L)$ and reflection $(R_R, R_L)$ coefficients of the considered structures, which depends on the material properties and geometry parameters. The detail of the TMM is provided in the method section.

For 'on demand' control of the reflection properties, we need the optimal complex permittivity distribution for every given structure. We require that the product of left and right reflections $\sqrt{R_R R_L}$ vanishes; a simple condition satisfied when at least one of the reflections diminishes while the other one remain non-zero, while keeping perfect transmission[4], $T_R = T_L = 1$. In this article, we design structures with purposefully asymmetric and frequency dependent reflections. In particular, a genetic algorithm streamlines the design process of unidirectional reflectionless structures starting from for a given (arbitrary) real permittivity distribution. The genetic optimization searches for the optimized parameters by minimizing a predefined target function through processes which tend to mimic the process of natural evolution. The target function, $F$, can be defined in different possible ways depending on the requirements. In our case, the target function is a measure of broadband unidirectionality i.e. perfect transmission with zero reflection from the left side, $R_L = 0$, that can be achieved with the following frequency dependent target function:

$$F(\omega) = \frac{1}{N} \int [|1 - T_R(\omega)| + |1 - T_L(\omega)| + |A(\omega)|] d\omega \qquad (1)$$

where $A(\omega)$ is the measure of an asymmetry in the reflectivity.

**Asymmetric relectivity in periodic structures**. We start from a simple three-layer periodic structure with given real part of permittivity values, and search for the corresponding imaginary parts to show the frequency selective unidirectional behavior such that $A(\omega) = 2R_L(\omega)/[R_L(\omega) + R_R(\omega)]$. Then, we increase the number of layers in the system and find the optimized values for unidirectional light propagation for desired range of frequencies. The genetic algorithm efficiently optimizes the permittivity distributions for asymmetric light effects in arbitrary layered structures. The scattering properties of optimized three, five and seven-layer systems for given real permittivity values are shown in Fig. 2. In all cases, the optimized imaginary parts permit perfect transmission from both sides while the reflection of left incident waves goes to zero, as anticipated in the target function. The spatial spectra of the optimized structure show an asymmetric coupling between wave vectors, responsible for left /right light propagation. The diminishing of spatial negative frequency components in the spectra, shown in right panels, confirms the asymmetric propagation behavior. As it is evident from the periodic nature of the proposed structure, the permittivity values form a closed loop in the complex index plane. Note that the asymmetry sign and the direction of the closed loop can be changed to clockwise by reversing the sign of imaginary parts of the permittivity and equivalently allowing reflectionless behavior for right incident waves, $R_R = 0$. Precisely, such enclosed area in the complex plane of the calculated permittivity profile provides an insight of the asymmetric response of the structure, which is ultimately proportional to $(R_R - R_L)$, for further details, see methods section.

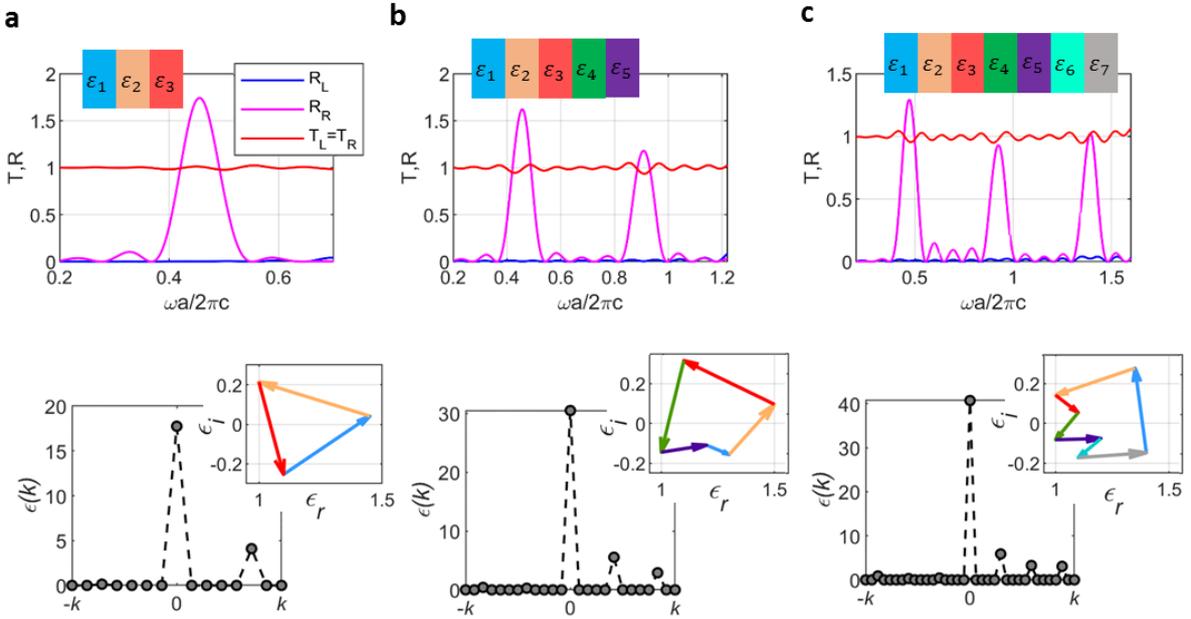

**Fig.2** Scattering properties of the optimized periodic non-Hermitian layer structure. **a** three-layer, **b** five-layer and **c** seven-layer periodic structure optimized with genetic algorithm for different real permittivity distribution seed. (top row) Transmission and reflection coefficients for the right and left illumination directions. (bottom row) Spectra of the optimized complex permittivity distribution showing the maximum asymmetry that induce unidirectional behavior. The insets illustrate the optimized permittivity values form a closed loop in complex plane. The unidirectionality occurs at around frequencies corresponding to the period of the structure $\omega a/2\pi c \approx 0.45$, and at the higher harmonics of that resonant frequency.

Next, we design non-Hermitian structures with unidirectional propagation directions depending on the operating frequency. As a particular example, we design a unidirectional reflectionless propagation in the forward direction for an operating frequency $\omega$ switched to backward direction by changing the operating frequency to $2\omega$. We define the target function the same as in Eq. (1) but to achieve the desired frequency dependent reflectivity, $A(\omega)$, in the following form: $A(\omega) = H(\omega - \omega_0)A_L(\omega) + H(\omega - \omega_0)A_R(\omega)$ where the terms $A_L(\omega) = 2R_L(\omega)/[R_L(\omega) + R_R(\omega)]$ and $A_R(\omega) = 2R_R(\omega)/[R_L(\omega) + R_R(\omega)]$ represent the measure of asymmetry in left and right reflectivity, respectively and $H(\omega)$ is the step function whose value is zero (one) for normalized frequencies $\omega a/2\pi c < 0.7$ and one (zero) for $\omega a/2\pi c \geq 0.7$ to switch the unidirectional behavior from left to right (right to left) with the operating frequency, $\omega_0$. To demonstrate such frequency dependent left/right unidirectional effect, we consider a five layer periodic structure with given real permittivity values and determine the imaginary part of permittivity in each layer such that the system shows the right unidirectional propagation for $\omega a/2\pi c \approx 0.45$ and left unidirectional propagation for $\omega a/2\pi c \approx 0.9$ and vice versa, as depicted in Fig. 3. The asymmetry between different wave vectors in the calculated spectra clearly shows that the designed structures exhibit frequency dependent unidirectional left/right propagation.

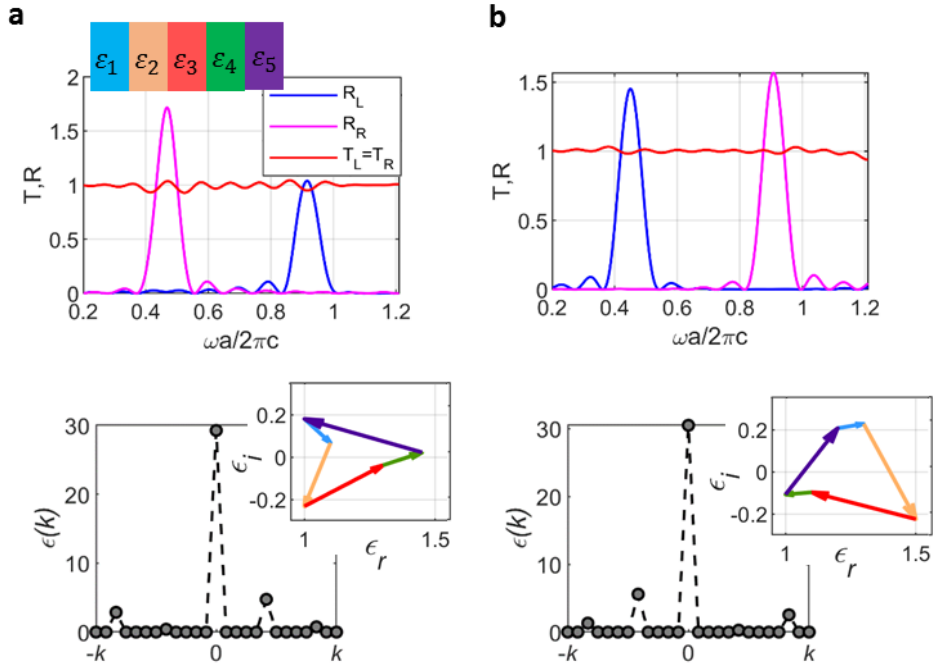

**Fig.3** Scattering properties of the optimized five-layer periodic system for frequency selective unidirectional behavior. The transmissions and reflections are obtained with genetic algorithm with different target functions and different real permittivity distribution seed. **a** right unidirectional propagation for $\omega a/2\pi c \approx 0.45$ and left unidirectional propagation for the $\omega a/2\pi c \approx 0.9$ **b** vice versa. (top row) Transmission and reflection coefficients for the right and left illumination directions. (bottom row) Spectra of the complex permittivity distribution for optimized structure showing the maximum asymmetry between different wave vectors that induce unidirectional behavior. The insets illustrate the optimized permittivity values, forming a closed loop, in complex plane

**Asymmetric reflectivity in nonperiodic structures**. Finally, we design the non-Hermitian structures that showing omnidirectional reflectionless propagation for a broad range of frequencies. Such structures rely on the generalized Hilbert transform that can be implemented with sophisticated meta-atoms[31,32]. Here, we propose a simple design to realize generalized HT with planar-layered structure for arbitrary permittivity distributions. We consider a non-periodic structure in which the central layers have arbitrary real permittivity values while the lateral layers on both left and right sides contain constant real part, for instance air, as depicted in Fig. 4. The structure is modified with an appropriate choice of gain-loss regions that suppress reflection in left direction to achieve unidirectional reflectionless effect. The target function is the same as that in Eq. (1) to eliminate the reflection of the left incident waves. The results for two different real permittivity distributions are shown in Fig. 4b and 4c. The determined imaginary permittivity values with genetic optimization confirms the unidirectional behavior by suppressing the reflection from the left side along with perfect transmission [see Fig. 4a(iii) and 4b(iii)]. The omnidirectional reflectionless behavior can be noticed in respective spectra where the spatial frequencies contributing to backward propagation are eliminated [see Fig. 4b(iv) and 4c(iv)].

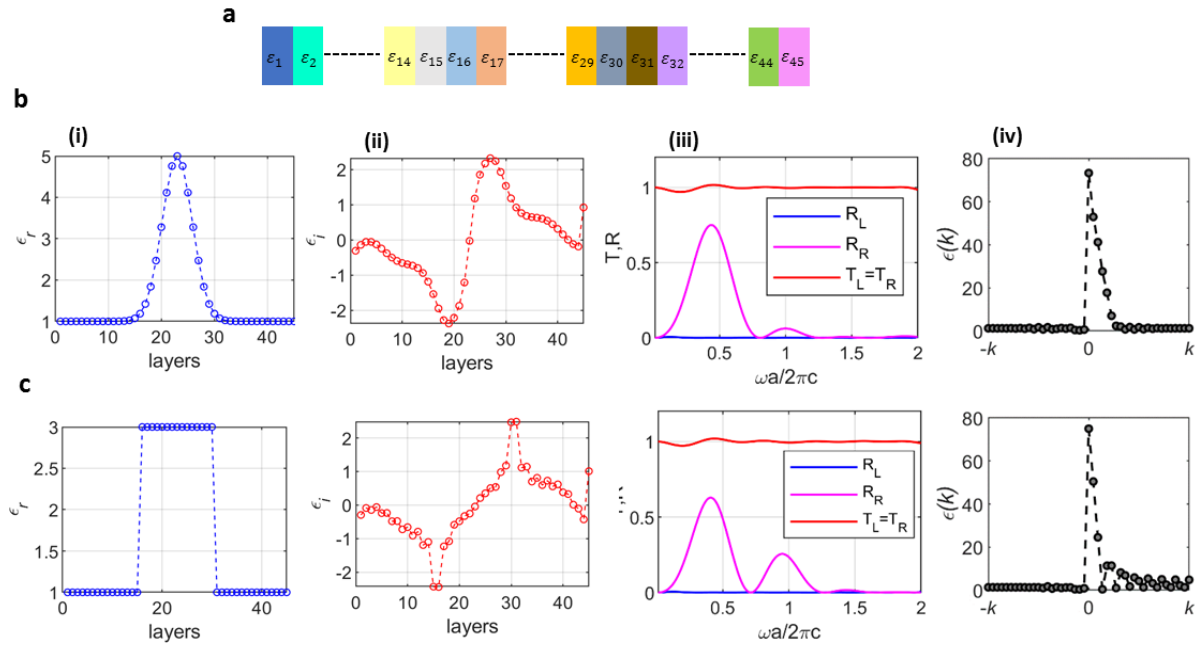

**Fig.4** Scattering properties of the designed non-periodic non-Hermitian layer structure. **a** Schematic illustration of the non-periodic structure. Two different distribution of real permittivity are considered to design unidirectional propagation for all spatial frequencies. **b** Gaussian shaped and **c** square shaped profiles of the seed. In both cases, (i) real parts of given permittivity values in each layer (ii) imaginary parts of corresponding permittivity values determined from genetic optimization (iii) transmission and reflection coefficients for unidirectional propagation (iv) Spectra of the complex permittivity distribution showing the maximum asymmetry among all spatial frequencies that allow unidirectional behavior.

**Conclusions:** We propose a new approach relying on genetic algorithm, as an intelligent engine to realize 'on demand' manipulation of light in non-Hermitian systems. In particular, we design periodic and non-periodic non-Hermitian structures for asymmetric light propagation in broad frequency range. We demonstrate that the procedure allows designing frequency dependent reflectionless structures for different directionalities with optimal complex permittivity distribution. The resulting asymmetric behavior of the designed structure is directly related to the enclosed area of the calculated permittivity distribution in complex space. The spectral properties confirm asymmetric reflectivity in the desired direction and frequencies ranges. Moreover, the proposed design recipe in the optical domain can also be extended to other classical wave systems, such as microwave, acoustics and elastic waves to achieve 'on demand' unidirectional light transport.

## Methods

**Transfer Matrix Method (TMM):**

TMM has been used to determine the scattering properties of the considered system, which is a common method describing the interaction between forward and backward propagating waves in multilayered structures with coherent interfaces. The total field, outside the scattering domain, may be expressed as the sum of the forward and backward propagating waves with wavevector, $k$, as: $E^-(z) = E_f^- e^{-ikz} + E_b^- e^{ikz}$ and $E^+(z) = E_f^+ e^{ikz} + E_b^+ e^{-ikz}$ for left $(z < -L)$ and right $(z > L)$ side of the structure, respectively. The interaction of waves at the different interfaces forms the transfer matrix, $M$, which couples the amplitudes of the forward and backward propagating waves at the left and right side of the structure:

$$\begin{pmatrix} E_f^+ \\ E_b^+ \end{pmatrix} = M \begin{pmatrix} E_f^- \\ E_b^- \end{pmatrix}, \quad M = \begin{pmatrix} M_{11} & M_{12} \\ M_{21} & M_{22} \end{pmatrix}, \tag{2}$$

Applying the boundary condition $E_b^+ = 0$ or $E_b^- = 0$ for either left ($L$) or right ($R$) incident waves respectively, we obtain the transmission coefficient $t_{R,L}$, and reflection coefficient $r_{R,L}$, (along with the transmittance $T_{R,L} = |t_{R,L}|^2$ and reflectance $R_{R,L} = |r_{R,L}|^2$), from the components of transfer matrix in the following form:

$$t_R = \frac{1}{M_{22}}, \quad t_L = \frac{M_{11}M_{22} - M_{12}M_{21}}{M_{22}}$$

$$r_R = \frac{M_{12}}{M_{22}}, \quad r_L = -\frac{M_{21}}{M_{22}} \tag{3}$$

**Asymmetric response of the structure:**

The asymmetric response of the structure, i.e the left-right reflection asymmetry, $(R_R - R_L)$, can also be analytically assessed. We start from the 1D Helmholtz equation:

$$\frac{\partial^2 E(z)}{\partial z^2} + k^2 \varepsilon(z) E(z) = 0 \qquad (4)$$

and a field composed by forward and backward waves: $E(z) = E_f(z)e^{ikz} + E_b(z)e^{-ikz}$. Under first Born approximation, the right to left reflection coefficient for the field amplitude is simply: $r_L(k) = \frac{ik}{2}\int_{-\infty}^{+\infty} e^{2ikz}\varepsilon(z)dz$, and more significantly the intensity reflection coefficient can be expressed as $R_R(k) = \frac{k^2}{4}\iint e^{2ik(z_1-z_2)}\varepsilon(z_1)\varepsilon^*(z_2)dz_1 dz_2$. The variable change $z_{\pm} = (z_1 \pm z_2)/2$ allows to write it as: $R_R(k) = \int \frac{1}{2}|\varepsilon(z_+)|^2 dz_+ \int k^2 e^{4ikz_-}dz_- + \int \frac{1}{2}(\varepsilon(z_+)d\varepsilon^*(z_+)/dz_+ - \varepsilon^*(z_+)d\varepsilon(z_+)/dz_+)dz_+ \int k^2 e^{4ikz_-}dz_-$, assuming that $|z_-|$ is small. Considering a periodic structure, the last expression integrated in one period can be written in the form: $R_R(k) = (\int \frac{1}{2}|\varepsilon(z_+)|^2 dz_+ + \int d\varepsilon_r d\varepsilon_i)\int k^2 e^{4ikz_-}dz_-$, where the second term inside brackets directly involves the area enclosed in the complex plane of profile permittivity: $\varepsilon(z_+) = \varepsilon_r(z_+) + i\varepsilon_i(z_+)$, when it is integrated along $z_+$ in one period. The sign of this term changes when considering the left to right reflection $R_L(k) = (\int \frac{1}{2}|\varepsilon(z_+)|^2 dz_+ - \int d\varepsilon_r d\varepsilon_i)\int k^2 e^{4ikz_-}dz_-$ and, thus, the reflection asymmetry is directly proportional to this area: $R_R(k) - R_L(k) \propto \int d\varepsilon_r d\varepsilon_i$.

**Acknowledgements**

The work described in here is partially supported by King Abdullah University of Science and Technology (KAUST) Office of Sponsored Research (OSR) under Award No. OSR-2016-CRG5-2950, KAUST Baseline Research Fund BAS/1/1626-01-01 and by NATO SPS research grant No: 985048. K.S. acknowledges funding from European Social Fund (project No 09.3.3-LMT-K712-17- 0016) under grant agreement with the Research Council of Lithuania (LMTLT).